\documentclass[twocolumn,amsmath,amssymb,aps,pra,longbibliography,floatfix]{revtex4-1}
\usepackage{physics}
\usepackage{units}
\usepackage{amsmath}
\usepackage{cases}
\usepackage{url}
\usepackage{natbib}
\usepackage{textcase}
\usepackage{amssymb}
\usepackage{graphicx}
\usepackage{bm} 

\usepackage{times}
\usepackage{float}
\usepackage{multirow,microtype,color,relsize}
\usepackage{subfigure}
\usepackage[breaklinks=true]{hyperref}
\usepackage[utf8]{inputenc}
\usepackage[english]{babel}
\hypersetup{colorlinks=true,linkcolor=blue,citecolor=blue}
\hypersetup{linktocpage}
\usepackage{CJKutf8}
\usepackage{breakcites}
\usepackage{float}
\usepackage{siunitx}
\usepackage[dvipsnames]{xcolor}
\definecolor{mypine}{RGB}{1, 121, 111}

\begin{document}
\begin{CJK*}{UTF8}{gbsn}
\title{Disorder effects in topological insulator nanowires}

\author{Yi Huang~(黄奕)}

\email[Corresponding author: ]{huan1756@umn.edu}
\author{B.\,I. Shklovskii}
 
\affiliation{School of Physics and Astronomy, University of Minnesota, Minneapolis, Minnesota 55455, USA}
\date{\today}

\begin{abstract}
Three-dimensional topological insulator (TI) nanowires with quantized surface subband spectra are studied as a main component of Majorana bound states (MBS) devices. However, such wires are known to have large concentration  $N \sim 10^{19}$ cm$^{-3}$ of Coulomb impurities. 
It is believed that a MBS device can function only if the amplitude of long-range fluctuations of the random Coulomb potential $\Gamma$ is smaller than the subband gap $\Delta$. 
Here we calculate $\Gamma$ for recently experimentally studied large-dielectric-constant (Bi$_{1-x}$Sb$_x$)$_2$Te$_{3}$ wires in a small-dielectric-constant environment (no superconductor). 
We show that provided by such a dielectric-constant contrast, the confinement of electric field of impurities within the wire allows more distant impurities to contribute into $\Gamma$, leading to $\Gamma \sim 3\Delta$. 
We also calculate a TI wire resistance as a function of the Fermi level and carrier concentration due to scattering on Coulomb and neutral impurities, and do not find observable discrete subband-spectrum related oscillations at $N \gtrsim 10^{18}$ cm$^{-3}$. 
\end{abstract}
\maketitle
\end{CJK*}

\section{Introduction}

Topological insulator (TI) attracts strong attention because of its conducting surface states. 
TI nanowires are predicted to host Majorana bound states (MBS) on their ends~\cite{fu2008,hasan2010,cook2011,cook2012,alicea2012,legg2021}.
Here we study the role of Coulomb impurities in TI nanowires. 

Typically, as-grown TI crystals are heavily doped semiconductors with concentration of donors $N \sim 10^{19}$ cm$^{-3}$. 
(For certainty, we talk about n-type case where the Fermi level $E_F$ is high in the conduction band.)
However, to make the bulk insulating and employ the surface states in transport, one has to move the Fermi level close to the Dirac point. 
This is done by intentional compensation of donors with an almost equal concentration of acceptors. 

This seemingly easy solution of the Fermi-level problem, however, comes with a price~\cite{skinner2012}. 
In fully compensated TI, all donors and acceptors are charged, and these randomly distributed charges in space create random potential fluctuations as large as the TI bulk semiconductor gap. 
These fluctuations create equal numbers of bulk electron and hole puddles, and substantially reduce the activation energy of the bulk transport~\cite{skinner2012,ren2011,knispel2017}.
At the same time near the surface, the random potential of charged impurities creates surface electron and hole puddles and smears the Dirac point by the energy $\Gamma$ self-consistently determined by the surface electrons (holes) screening~\cite{skinner2013a,skinner2013b}. This smearing was observed by the scanning tunnel microscopy~\cite{beidenkopf2011}. 

Similar Dirac point smearing by disorder happens at the surface of thin TI films sandwiched between two low-dielectric-constant layers~\cite{huang2021a}, where the Rytova-Chaplik-Entin-Keldysh modification~\cite{rytova1967,chaplik1971,keldysh1979} of the Coulomb potential of a charge impurity slows down the potential decay in space and allows a larger number of the film impurities to contribute in $\Gamma$. In such films there is no bulk puddles because of strong screening of surface electrons.  

Recently, substantial experimental progress has been made in thin TI nanowires and nanoribbons, where all mobile electrons (holes) are located at the wire sufrace ~\cite{jauregui2016,arango2016,ziegler2018,munning2021,peng2010,hong2012,hamdou2013,cho2015,rosenbach2020,rosenbach2021,breunig2021}.
They attract attention because the Dirac spectrum of surface electrons (holes) splits in equidistant surface subbands separated by the substantial gap $\Delta$ due to the small wire cross-section perimeter. 
In such a TI wire, one can realize MBS with the help of superconductor proximity effect, if one is able to keep the Fermi level within a certain, say the first, subband in the whole length of the TI wire~\cite{legg2021}. 
For such a fine tuning, the subband gap $\Delta$ should be bigger than the amplitude $\Gamma$ of fluctuations of the long-range electrostatic potential along the wire due to charged impurities. 
In this case the resistance $R(E_F)$ should oscillate with the Fermi level $E_F$. 
Reference~\cite{munning2021} claims small resistance oscillations in TI wires of radius $a=20$ nm on relatively thick silicon oxide substrate separating it from the heavily doped Si gate in the absence of any superconductor. These oscillations seemingly indicate that $\Gamma \lesssim \Delta$.

In this paper, we calculate $\Gamma$ for a cylindrical TI wire of the radius $a$ with concentration $N$ of Coulomb impurities in the set up of Ref.~\cite{munning2021}. To this end we study the self-consistent screening of Coulomb impurities by surface electrons and holes, in a way similar to Refs.~\cite{skinner2013a,huang2021a}. 
We show that $\Gamma \sim 3\Delta$ for $N \sim 10^{19}$ cm$^{-3}$. 
We also calculate a TI wire resistance $R(E_F)$ due to scattering on the Coulomb and short-range impurities and find no observable oscillation at $N \gtrsim 10^{18}$ cm$^{-3}$ because $\Gamma \gtrsim \Delta$. 

Our results may seem counterintuitive. 
One might think that the role of disorder in a thin TI wire should be smaller than in the bulk TI or in a TI film. 
Indeed, at a given three dimensional (3D) concentration of charged impurities $N$, the 1D concentration of them $N \pi a^2$ in a thin TI wire is quite small, so that there are less impurities at a given distance $r > a$ from an observation point in this case.
We, however, show that this larger-distance effect is compensated by the effect of confinement of the electric field inside the TI wire due to its large dielectric constant $\kappa \sim 200$. 
As a result, the Coulomb potential of each impurity slowly decays along a wire. 
This allows a larger number of Coulomb impurities to contribute to $\Gamma$. 

The plan of this paper is as follows. In Sec. II we calculate the amplitude $\Gamma$  of long-range potential fluctuations produced by randomly positioned Coulomb impurities inside the wire. 
In Sec. III we calculate a TI wire resistance limited by the scattering of surface electrons on bulk charged impurities. 
In Sec. IV we calculate additional contribution to the resistance from located at the surface point-like neutral impurities. 
We conclude in Sec. V by comments on the role of a superconductor stripe attached to the surface of the studied above TI wire and on the requirement for the concentration of Coulomb impurities $N$ for the proper functioning of a resulting MBS device.

\section{self-consistent calculation of disorder potential}

\begin{figure}[t]
    \centering
    \includegraphics[width=\linewidth]{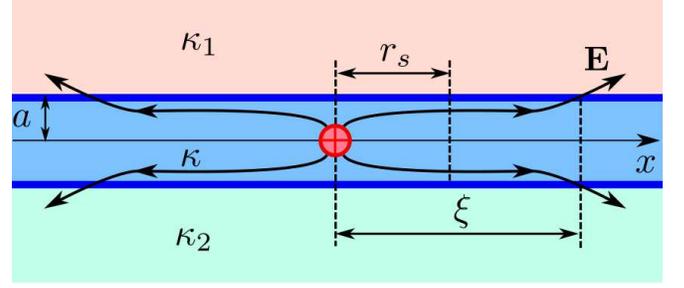}
    \caption{Schematic plot of a wire of radius $a$ and dielectric constant $\kappa$. The dielectric constants for the environment are $\kappa_1$ on the top and $\kappa_2$ on the bottom of the wire, with an average $\kappa_e = (\kappa_1 + \kappa_2)/2$. In a topologically trivial case with $\kappa \gg \kappa_e$ electric field lines (black) start at the charged impurity shown by a red circle and channel through the wire until exiting outside at $x\sim \xi \gg a$. In a TI wire with $\kappa \gg \kappa_e$ the electric field lines are screened (not shown) at distance $r_s < \xi$ by the surface electrons or holes (heavy blue). }
    \label{fig:wire}
\end{figure}

Consider the electric field of a Coulomb impurity located inside an infinitely long ($L_x \to \infty$, no surface electrons or holes) trivial wire of radius $a$ and with dielectric constant $\kappa$. 
The wire is in an environment with a dielectric constant $\kappa_e$ such that $\kappa \gg \kappa_e$ (see Fig.~\ref{fig:wire}). 
In this case, due to large dielectric-constant mismatch, the electric field lines of a point charge are trapped inside the wire for a long distance $\xi \gg a$ before exiting outside. 
At distances $\xi \gg \abs{x} \gg a$, using Gauss theorem, this uniform electric field can be found to be $E_0 = 2e/\kappa a^2$, and $e$ is the proton charge. 
More accurately, the electrostatics of this problem for a topologically trivial wire was solved and used in Refs.~\cite{keldysh1997,finkelstein2002,teber2005,kamenev2006,cui2006}, and below we summarize their results.
The characteristic length $\xi$ is determined by the equation $\xi^2 = a^2 (\kappa/2\kappa_e) \ln(2 \xi /a)$. 
The electric field and the Coulomb potential at $a< \abs{x} < \xi \ln(\kappa/\kappa_e)$ decay exponentially:
\begin{align}\label{eq:Ex_exp}
    E(x) &= E_0 e^{-\abs{x}/\xi}, \\
    v_0(x) &= e E_0 \xi e^{-\abs{x}/\xi}, \label{eq:phix_exp}
\end{align}
with the distance $x$ away from the impurity center.
This decay is replaced by the conventional isotropic Coulomb asymptotic $E(r) = e/\kappa_e r^2$ and $v_0(r) = e^2/\kappa_e r$ at $r \gtrsim \xi \ln(\kappa/\kappa_e) \gg \xi$. 
In Eq.~\eqref{eq:phix_exp} and below we ignore a constant term smaller than $e E_0 \xi$ which comes from distances $r > \xi$.
 
The Fourier transform of Eq.~\eqref{eq:phix_exp} is given by
\begin{align}\label{eq:v0q}
    v_0(q) = \int \limits_{-\infty}^{\infty} dx \, v_0(x) e^{-i qx} = \frac{2eE_0}{q^2 + \xi^{-2}}.
\end{align}
where $q < a^{-1}$ since $\abs{x} > a$.

In the absence of free electrons (holes) screening, Eq.~\eqref{eq:phix_exp} leads to the amplitude of potential fluctuations of charged impurities $\sim eE_0 \xi \sqrt{N \pi a^2 \xi}$. 
For highly compensated (Bi$_{1-x}$Sb$_x$)$_2$Te$_{3}$ (BST) nanowires used in experiments~\cite{munning2021} with $a = 20$ nm, $N = 10^{19}$ cm$^{-3}$, $\kappa = 200$~\cite{richter1977,borgwardt2016,bomerich2017}, and $\kappa_e = 2.5$ (average of the vacuum $\kappa_1=1$ and silicon oxide values $\kappa_2 = 4$), we have $\xi = 11 a$ and this potential fluctuation is of order $0.4$ eV.

To describe the screening by surface Dirac electrons and holes, we start from the equation for the electric potential of screened charged impurities $\phi(x)$
\begin{equation}
   \mu [\lambda(x)] - e \phi(x) = E_F,
\end{equation}
where $E_F$ is the (gate tuned) Fermi level (electro-chemical potential), $\mu[\lambda(x)]$ is the (local) chemical potential, and $-e\lambda(x)$ is the (local) linear charge density of carriers.
Explicitly, we can calculate $\lambda$ as a function of $\mu$ through 
\begin{align}
    \lambda(\mu) = &\frac{{\rm sgn}(\mu)}{L_x}  \sum_{k,l} \Theta(\mu^2 - \varepsilon_l^2(k)) \nonumber \\
    = &\frac{{\rm sgn}(\mu)}{\pi\hbar v_F} \sum_{l} \Theta(\mu^2 - \varepsilon_l^2) \sqrt{\mu^2 - \varepsilon_l^2} , \label{eq:lambda}
\end{align}
where ${\rm sgn}(\mu)$ is the sign function, $\Theta(x)$ is the Heaviside theta function, and the energy spectrum for Dirac electrons is given by
\begin{align}
    \varepsilon_l(k) = \hbar v_F \sqrt{k^2 + (l/a)^2},
\end{align}
and Dirac holes spectrum is $-\varepsilon_l(k)$.
Here $v_F$ is the Fermi velocity, $k$ is the momentum along wire, and $l = \pm 1/2, \,\pm 3/2, \,\dots$ are angular momentum quantum number due to antiperiodic boundary condition. 
At $k=0$, neighboring subbands are separated by equidistance gap $\Delta = \hbar v_F/a$.
Here in Eq.~\eqref{eq:lambda} and below we abbreviate $\varepsilon_l(k=0) = \abs{l} \Delta $ as $\varepsilon_{l}$ for convenience.
In experiments with BST TI nanowires with radius $a=20$ nm~\cite{munning2021}, $\Delta \sim 13$ meV which, as we will see below, is much smaller than the amplitude of fluctuations of screened Coulomb potential $\Gamma \sim 35$ meV. 
Therefore, it is natural to consider the limit $\mu^2 \gg \Delta^2$, such that the sum over $l$ in Eq.~\eqref{eq:lambda} can be approximated by an integral and gives 
\begin{align}
    \lambda(\mu) = {\rm sgn}(\mu) \frac{a}{2} \qty(\frac{\mu}{\hbar v_F})^2,
\end{align}
which is related to the 2D charge concentration of the wire surface $n = \lambda/2\pi a = k_F^2/4\pi$, where $k_F = \mu/\hbar v_F$ is the (local) 2D Fermi wave vector.

If $E_F$ is large enough so that $E_F^2 \gg e^2\phi^2$, then $\mu[\lambda(x)]$ can be linearized in the local charge concentration variation $\delta \lambda (x)$,
\begin{align}
    \mu[\lambda(x)] = E_F + \delta \lambda (x)/g(\mu).
\end{align}
where $g(\mu) = d{\lambda}/d{\mu}$ is the density of states (DOS) at zero temperature.
Explicitly using Eq.~\eqref{eq:lambda} we have 
\begin{align}\label{eq:tdos}
    g(\mu) = \frac{\kappa \alpha \abs{\mu}}{\pi e^2} \sum_{l} \frac{\Theta(\mu^2 - \varepsilon_l^2)}{\sqrt{\mu^2 - \varepsilon_l^2}}, 
\end{align}
where $\alpha = e^2 / \kappa \hbar v_F$ is the effective fine structure constant.
In continuous limit $\mu^2 \gg \Delta^2$, the DOS has a simple expression
\begin{align}\label{eq:dos_g_cont}
    g(\mu) = \frac{\abs{\mu}}{a \Delta^2},
\end{align}
which also reminds us of the 2D DOS $\nu(\mu) = g(\mu)/2\pi a$.

In the Thomas-Fermi (TF) approximation the dielectric function is given by
\begin{equation}
    \epsilon(q) = 1- v_0(q) \Pi_{TF},
\end{equation}
where the TF polarization bubble is $\Pi_{TF} = -g(\mu)$, and the bare interaction $v_0(q)$ is given by Eq.~\eqref{eq:v0q}.
We arrive at the screened potential of one charge impurity within a thin TI wire,
\begin{equation}\label{eq:vq}
    v(q) = \frac{v_0(q)}{\epsilon(q)} = \frac{2eE_0}{q^2 + r_s^{-2}},
\end{equation}
where 
\begin{align}\label{eq:rs}
    r_s = [\xi^{-2} + 2eE_0 g(\mu)]^{-1/2}
\end{align}
is the screening length and $q < a^{-1}$.
In real space, the screened potential 
\begin{align}\label{eq:vx_screened}
    v(x) = eE_0 r_s e^{-\abs{x}/r_s}
\end{align}
is similar to the bare potential Eq.~\eqref{eq:phix_exp}, but with a different characteristic length $r_s < \xi$.

TF approximation used above for calculation of $r_s$ is valid if $r_s$ is much larger than the Fermi wavelength $\sim k_F^{-1}$. 
We show below that this condition is well satisfied.

If impurities are randomly distributed inside the wire, the mean-squared fluctuation of the potential is given
by
\begin{align} 
\label{eq:phi2}
\ev{\phi^2} &= \frac{1}{e^2}\int \limits_{-\infty}^{\infty}  dx\; v^2(x) N\pi a^2 = N \pi a^2 E_0^2 r_s^3.
\end{align}

In the rest of this section, we concentrate on the charge neutrality point where $E_F = 0$, and $\phi$ has the Gaussian distribution function with $\ev{\phi} = 0$ and $\ev{\phi^2} = \Gamma^2/e^2$. 
Next we calculate the average DOS $\ev{g}$ using the Gaussian distribution function of $\phi$:
\begin{align}
    \ev{g} &= \int_{-\infty}^{\infty} d(e\phi) g(e\phi) \frac{e^{-e^2\phi^2/2\Gamma^2}}{\sqrt{2\pi \Gamma^2}}\\ \nonumber
    &= \frac{\kappa \alpha}{\pi e^2} \sum_{l} \exp(-\varepsilon_l^2/2\Gamma^2).
\end{align}
In the limit $\Gamma \gg \Delta$, the summation over $l$ can be replaced by an integral and we arrive at 
\begin{align}
    \ev{g} &= \sqrt{\frac{2}{\pi}} \frac{\Gamma}{a \Delta^2} .
\end{align}
The same result can be obtained if we average Eq.~\eqref{eq:dos_g_cont} directly
~\footnote{In the opposite limit $\Gamma \ll \Delta$, the average DOS is exponentially suppressed by the gap and leads to $\ev{g} = (2\kappa \alpha/\pi e^2) e^{-\Delta^2/8\Gamma^2}$.}.
Replacing $g(\mu)$ by $\ev{g}$ and ignoring the first term in Eq.~\eqref{eq:rs} because as we will see $r_s \ll \xi$, we arrive at 
\begin{align}\label{eq:rs_scf}
    r_s \approx [2eE_0 \ev{g}]^{-1/2} = \frac{a}{2} \qty(\frac{\pi}{2})^{1/4}  \qty(\frac{\Delta}{\alpha \Gamma})^{1/2}.
\end{align}
Substituting this $r_s$ back into Eq.~\eqref{eq:phi2} and using $\Gamma^2 = e^2 \ev{\phi^2}$, we can solve for $\Gamma$ 
\begin{align}
    \Gamma = \qty(\frac{\pi}{2})^{1/2} \alpha^{1/7} (Na^3)^{2/7} \Delta
    = \qty(\frac{\pi}{2})^{1/2} \frac{e^2}{\kappa a} \frac{(Na^3)^{2/7}}{\alpha^{6/7}} . \label{eq:gamma}
\end{align}
Using $a = 20$ nm, $\kappa = 200$, and $N = 10^{19}$ cm$^{-3}$, 
$v_F = 4 \times 10^5$ m/s~\cite{jszhang2011}, one gets $\alpha = 0.027$, $\Gamma = 35$ meV and $\Delta = 13$ meV.
Thus, our assumption $\Gamma \gg \Delta$ is well justified.

The TF approximation is justified as well, because the Fermi wavelength $\sim k_F^{-1} = \hbar v_F/\Gamma$ is much smaller than the screening length $r_s$ given by Eq.~\eqref{eq:rs_scf}.
Indeed, the ratio $1/k_F r_s \sim (\alpha \Delta/\Gamma)^{1/2} \ll 1$ given $\Gamma \gg \Delta$ and $\alpha \ll 1$, so TF approximation is justified in our case. 

Note that according to Eq.~\eqref{eq:gamma} $\Gamma$ does not decrease with decreasing $a$ in spite of the decrease of linear concentration $\pi a^2 N$ of Coulomb impurities. 
This is a result of the confinement of electric field lines inside the large-dielectric-constant TI wire described in Eq.~\eqref{eq:vx_screened}, which allows more distant impurities to contribute in the amplitude of the disorder potential. 

Substituting Eq.~\eqref{eq:gamma} back to Eq.~\eqref{eq:rs_scf}, we have
\begin{align} \label{eq:screening}
    r_s = \frac{a}{2} \alpha^{-4/7} (Na^3)^{-1/7}.
\end{align}
The above theory assumes that $r_s > a$, so that the random potential depends only on $x$ and we can avoid averaging of random potential over perimeter of a wire cross-section. Indeed, for $N = 10^{19}$ cm$^{-3}$ and $a = 20$ nm, we get $r_s = 2.1 a$, so that the above theory is (marginally) applicable. 
At the same time our assumption that $r_s \ll \xi=11 a$ is justified as well.

Thus, at $E_F=0$ the surface of a cylindrical TI wire is covered by alternating along the $x$ axis electron and hole puddles. Each puddle is a ring with radius $a$ and height $\Delta x \sim r_s$.
Each puddle contains electrons (holes) from different subbands and with both directions of $k$. This makes them spinless. Electron-electron correlations can make them spinful, but this does not affect our theory.


On the other hand, for larger $\alpha$ and $N$ such that $a > r_s$, i.e., $a > a_c = \alpha^{-4/3} N^{-1/3}$, one can neglect the curvature of the wire surface and our Eqs.~\eqref{eq:gamma} and \eqref{eq:screening} cross over to $\Gamma$ and $r_s$ for the bulk TI sample calculated in Ref.~\citep{skinner2013a}. 

Above we ignored the concentration of charged impurities in the environment outside of the TI wire $N_e$. 
Let us now evaluate the role of such impurities, say, in the silicon oxide substrate. 
In the case of trivial wire, in order to save electrostatic energy, the electric field lines of an impurity at distance $r < \xi$ from the wire surface first enter inside the wire and then spread inside the wire to the distance $\sim \xi$ before exiting outside the wire to infinity. 
Thus, one can think that effectively each outside impurity is represented inside the wire by a charge $e$ cylinder, with length $r$ and radius $a$. 
In the presence of surface screening of a TI wire, only a small minority of the outside impurities with $r < r_s$ contribute in fluctuating charge of the volume $a^2 r_s$. 
As a result, the total effective concentration of impurities projected from outside the wire is $N_e (r_s/a)^2$. 
If $N_e(r_s/a)^2\ll N$, where $r_s$ is given by Eq.~\eqref{eq:screening}, then impurities outside the wire can be ignored.
For example, for the BST TI wire with $N = 10^{19}$ cm$^{-3}$ and $a = 20$ nm on silicon oxide substrate with~\cite{shklovskii2007} $N_e \sim 10^{17}$ cm$^{-3}$, 
using $r_s/a\sim 2$, we get $N_e(r_s/a)^2 \sim 4\times 10^{17}$ cm$^{-3} \ll N$. 
Thus, outside impurities can be ignored~\footnote{On the other hand, if $N_e(r_s/a)^2> N$ the screening length $r_s$ should be recalculated self-consistently together with $\Gamma$. Then, instead of Eqs.~\eqref{eq:gamma} and \eqref{eq:screening}, we arrive at new results $\Gamma \sim \Delta \alpha^{-1/9} (N_e a^3)^{2/9}$ and $r_s \sim a \alpha^{-4/9} (N_e a^3)^{-1/9}$.}.

\section{Resistance due to Scattering on Coulomb impurities}
In this section, we calculate the resistance $R$ as a function of the Fermi level $E_F$ limited by the scattering off charged impurities.
Let us start with the conductivity of the TI wire surface. 
In the linear screening region $\mu^2 \gg e^2\phi^2$, where the carrier concentration is weakly perturbed by impurities, using Boltzmann kinetic equation for Dirac electrons (holes), one arrives at the surface conductivity~\cite{culcer2008},
\begin{equation}\label{eq:conduct}
    \sigma = \frac{e^2}{h} \frac{\abs{\mu} \tau}{4\hbar}.
\end{equation}
Here $\tau$ is the transport relaxation time whose inverse is given by~\cite{ando2006,adam2007,adam2009,bardarson2010}
\begin{equation}
    \frac{1}{\tau} = \frac{2\pi}{\hbar} \sum_{i,f} w_i \ev{\abs{V_{fi}}^2} \frac{1}{2}(1-\cos^2 \theta) \delta(\varepsilon_f - \varepsilon_i). \label{eq:rate}
\end{equation}
Here the subscripts $f(i)$ denote the final (initial) states with momentum $(k_f, l_f/a)$ [or $(k_i, l_i/a)$]; $w_i = g_{l_i}/g$ is the probability to be initialized in state $i$, where $g_{l}(\mu) = (\pi a \Delta)^{-1}\Theta(\mu^2 - \epsilon_l^2)\abs{\mu}/\sqrt{\mu^2 - \epsilon_l^2}$ is the DOS of the $l$th subband [c.f. Eq.~\eqref{eq:tdos}]; and $\ev{\dots}$ denotes the average over impurities position. 
The scattering angle $\theta$ can be related to the (local) transfer momentum $\vb{p} = (k_f-k_i, l_f/a - l_i/a)$ via $\cos \theta = 1 - \frac{p^2}{2k_F^2}$,
where $k_F = \mu / \hbar v_F$.
The extra factor $(1 + \cos \theta)/2$ in Eq.~\eqref{eq:rate} arises when the back scattering is suppressed as a consequence of the spin texture at the Dirac point, as in graphene or Weyl semimetals~\cite{ando2006,adam2007,adam2009,burkov2011}. 
In other words, because of spin-momentum locking near the Dirac point, the back scattering results in opposite orientations of spin. 
Since in our case of scattering there is no spin-flipping process, perfect back scattering is suppressed by spin-momentum locking.
However, it is still possible for an electron to change its propagation direction along the wire by scattering off impurities.
Namely, although for perfect back scattering $\theta = \pi$, which corresponds to $k_f = -k_i$ and $l_f = -l_i$, the scattering probability is zero, for non-perfect back scattering $\pi/2<\theta<\pi$ the scattering probability is reduced, but still is finite. 
In this sense, the surface states of TI nanowires are very different from the edge states in quantum Hall effect, for later the back scattering is completely forbidden when the distance between counter propagating edges is larger than the cyclotron radius.

Because the long-range Coulomb potential Eq.~\eqref{eq:vx_screened} is a function of $x$ only, the angular momentum is conserved and we have
\begin{align}\label{eq:long_range_vfi}
    \ev{\abs{V_{fi}}^2} = \delta_{l_i, l_f} \frac{N \pi a^2}{L_x} \abs{v(q)}^2,
\end{align}
where $\delta_{l_i, l_f}$ is the Kronecker delta, $L_x$ is the length of the wire, and $q = k_f - k_i$ is the momentum transfer along $x$.
Because of the angular momentum conservation $l_f = l_i$, we only need to consider intra-subband scattering, which greatly simplifies the calculation.
Since energy is also conserved, there are only two possibilities for the final momentum along the wire $k_f = \pm k_i$.
The scattering angle factor $(1 - \cos \theta)$ suppresses the contribution from the forward scattering $k_f = k_i$ and $l_f = l_i$, so the only remaining contribution comes from $k_f = - k_i$ and $l_f = l_i$.
Therefore, using Eqs.~\eqref{eq:vq}, \eqref{eq:rate}, and \eqref{eq:long_range_vfi}, we arrive at the rate 
\begin{align}
     \frac{1}{\tau(\mu)}  = \frac{1}{\tau_0(\mu)}  f(\mu). \label{eq:long_range_tau}
\end{align}
Here,
\begin{align}
    \frac{1}{\tau_0(\mu)} = \frac{16Na r_s^2 \alpha^2 \Delta^3}{\hbar \mu^2} = \frac{4Na^3 \alpha \Delta^4}{\hbar \abs{\mu}^3}\label{eq:long_range_tau_a},
\end{align}
[we used $r_s(\mu) = a \sqrt{\Delta/4\alpha \abs{\mu}}$ which follows from Eqs.~\eqref{eq:rs} and \eqref{eq:tdos} at $r_s \ll \xi$]
and the function 
\begin{align}
    f(\mu) = \frac{\alpha\Delta^2}{\mu^2}\sum_l \frac{\varepsilon_l^2}{\mu^2}\frac{\Theta(\mu^2 - \varepsilon_l^2)}{\qty(1-\frac{\varepsilon_l^2}{\mu^2} + \frac{a^2\Delta^2}{4r_s^2 \mu^2})^2} \label{eq:fmu}
\end{align}
oscillates as a function of $\mu$ around its average value $f=1$.
It peaks at half-integer $\mu/\Delta$ because of sharp peaks of the matrix element in Eq.~\eqref{eq:long_range_vfi} at small momentum transfer.
Below we use Eq.~\eqref{eq:long_range_tau_a} to study the monotonous dependence of the wire resistance on the Fermi energy $E_F$. 

Note that the right-hand side of Eq.~\eqref{eq:long_range_tau_a} is proportional to $r_s^2$ because the main contribution to the scattering rate is at small momentum transfer $q \sim r_s^{-1}$.
Such an unconventional role of long-range potential results from the electric field confinement in a wire which allows more distant Coulomb impurities to contribute into scattering~\footnote{Using $r_s(\mu) \approx a \sqrt{\Delta/4\alpha \abs{\mu}}$, we have $r_s > a$ if $\abs{\mu} < 9 \Delta$ which covers most of the range of interest in experiments.}.

At $\abs{E_F} \gg \Gamma$ and, therefore, $\abs{\mu} \simeq \abs{E_F} \gg \Gamma$ we can substitute Eq.~\eqref{eq:long_range_tau_a} into Eq.~\eqref{eq:conduct}. Substituting $E_F$ for $\mu$  we obtain for the wire resistance:
\begin{align}\label{eq:charged_rho}
    R(E_F)= \frac{L_x}{2\pi a\sigma} = \frac{h}{e^2} \frac{16 N a^3 \alpha}{(E_F/\Delta)^{4}}\frac{L_x}{2\pi a},
\end{align}

At $\abs{E_F} \lesssim \Gamma$, screening becomes nonlinear. 
Equation~\eqref{eq:charged_rho} saturates at $\abs{E_F} \simeq \Gamma$ and using Eq.~\eqref{eq:gamma} we obtain the maximum resistance
\begin{align}\label{eq:rho_max}
    R_{\max} \simeq \frac{h}{e^2} \frac{2^6 \alpha^{3/7}}{\pi^2 (Na^3)^{1/7}} \frac{L_x}{2\pi a}.
\end{align}
Now let us compare Eqs.~\eqref{eq:charged_rho} and ~\eqref{eq:rho_max} with experiment~\cite{munning2021}. 
Experimental resistance~\cite{munning2021} has a maximum $R_{\max} \sim 50$ k$\Omega$ at the Dirac point and slightly (by $\sim 10$\%) decreases at $\abs{E_F}/\Delta \simeq 9/2$. For $L_x = 1$ $\mu$m, the radius $a = 20$ nm, and $N=10^{19}$ cm$^{-3}$, Eq.~\eqref{eq:rho_max} predicts $R_{\max} \sim 150$ k$\Omega$. 
This maximum resistance is somewhat larger than the experimental one~\cite{munning2021}.
Our theory also predicts that the $R(E_F)$ stays constant only for $\abs{E_F} \lesssim \Gamma \simeq 3 \Delta$ and then decreases $ \propto E_F^{-4}$ which disagrees with experiments. 
This motivates us to explore in the next section whether the addition of surface short-range impurities to the bulk Coulomb ones can improve agreement with the experiment. 

\section{Contribution of short range impurities to a TI wire resistance}

The potential of short range impurities is $v(\vb{r}) = u_0 \sum_j \delta^{(2)}(\vb{r}-\vb{r}_j)$ where $\vb{r}_j$ is the 2D position of an impurity on the TI surface.
The corresponding matrix element is given by
\begin{align}
    \ev{\abs{V_{fi}}^2} = \frac{n_i u_0^2}{2\pi a L_x},
\end{align}
where $n_i$ is the 2D concentration of short range impurities on the wire surface.
In contrast to the case of long-range Coulomb impurities discussed in last section, both inter- and intrasubband scattering contribute in the scattering rate as shown in Eq.~\eqref{eq:rate}. 
Because the matrix element is independent on the subband indices, the summation over initial states is trivial $\sum_i w_i = 1$, while
summation over final states leads to the scattering rate
\begin{align}\label{eq:rate_short}
    \frac{1}{\tau(\mu)} = \frac{n_i u_0^2}{2\hbar a} g(\mu).
\end{align}
Substituting Eq.~\eqref{eq:rate_short} into Eq.~\eqref{eq:conduct}, we arrive at the contribution of short range impurities to the TI wire resistance 
\begin{align}\label{eq:rho_short}
    R(\mu) = \frac{h}{e^2} \frac{2 n_i u_0^2 g(\mu)}{a \abs{\mu}} \frac{L_x}{2\pi a}.
\end{align}
In continuous spectrum limit $\mu^2 \gg \Delta^2$, we replace $g(\mu) $ by Eq.~\eqref{eq:dos_g_cont} and obtain the independent-on-$E_F$ resistance
\begin{align}
    R_0 = \frac{h}{e^2} \frac{2 n_i u_0^2}{\hbar^2 v_F^2} \frac{L_x}{2\pi a}.
\end{align}
It was shown~\cite{munning2021} that, on the top of the above classical behavior, quantum oscillations of the DOS $g(\mu)$ [see Eq.~\eqref{eq:tdos}] result in resistance oscillations as a function of the Fermi level with period $\Delta$.
However, we show below that these oscillations are very small due to the smearing effect of Gaussian fluctuations of the long-range potential. 
For example, the amplitude of a harmonics with period $\Delta$ is suppressed by a factor $\exp[-2\pi^2\Gamma^2(E_F)/\Delta^2]$. 
The Fermi-level dependent $\Gamma(E_F)$ can be obtained by substituting Eq.~\eqref{eq:rs} into Eq.~\eqref{eq:phi2}, assuming $\xi \gg r_s$, and replacing $\abs{\mu}$ by $\sqrt{E_F^2 + 2\Gamma^2(E_F)/\pi}$:
\begin{align}\label{eq:gamma_mu}
    \Gamma(E_F) = \sqrt{\frac{\pi N a^3}{2}}  \frac{\alpha^{1/4} \Delta^{7/4}}{[E_F^2 + 2\Gamma^2(E_F)/\pi]^{3/8}}.
\end{align}
Indeed, by solving for $\Gamma = \Gamma(E_F = 0)$ self-consistently, we obtain the same expression $\Gamma$ given by Eq.~\eqref{eq:gamma}. 
At $\abs{E_F} \gg \Gamma$, one has $\Gamma(E_F) \propto \abs{E_F}^{-3/4}$.

\begin{figure}[t]
    \centering
    \includegraphics[width=\linewidth]{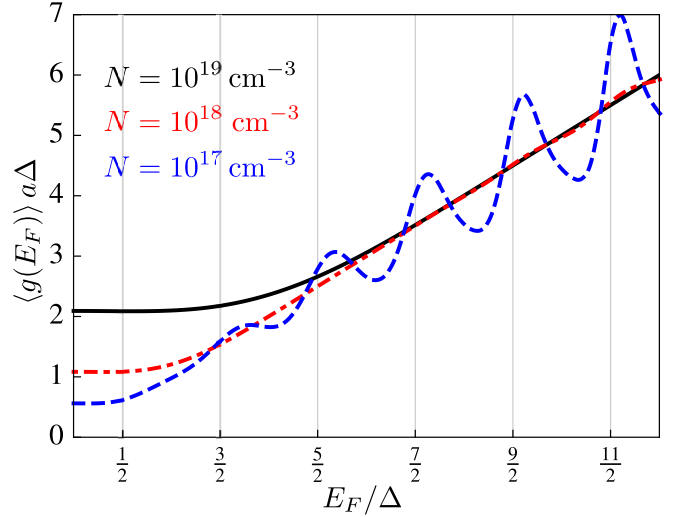}
    \caption{The DOS $\ev{g(E_F)}$ defined in Eq.~\eqref{eq:gmu_avg}, in units of $(a \Delta)^{-1}$, smeared by the long-range potential of Coulomb impurities with different 3D concentrations $N$ in units of cm$^{-3}$: $10^{17}$ (dashed blue line), $10^{18}$ (dot dashed red line), and $10^{19}$ (solid black line). Here we use radius $a = 20$ nm and $\alpha = 0.027$.}
    \label{fig:tdos}
\end{figure}
\begin{figure}[t]
    \centering
    \includegraphics[width=\linewidth]{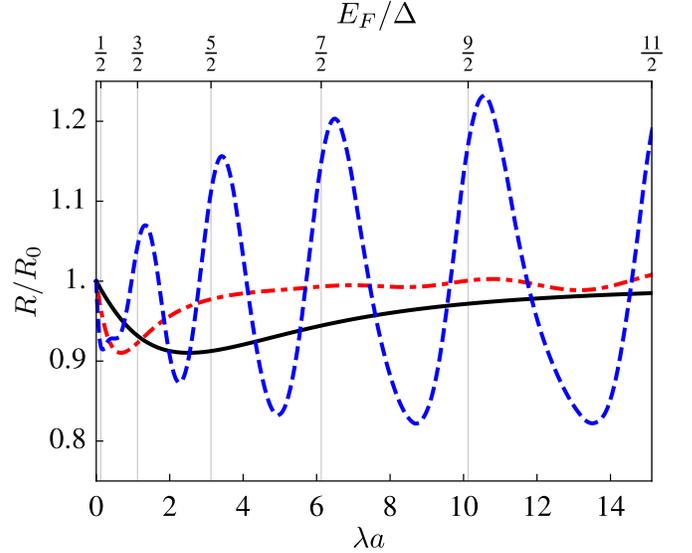}
    \caption{ Dimensionless resistance $R/R_0$ of a TI wire due to scattering on short-range surface impurities averaged over the long-range potential of Coulomb impurities as a function of $E_F/\Delta$ or of average 1D dimensionless charge concentration $\lambda a$. Different 3D concentrations $N$ of Coulomb impurities in units of cm$^{-3}$ are: $10^{17}$ (dashed blue line), $10^{18}$ (dot dashed red line), and $10^{19}$ (solid black line). Here we use radius $a = 20$ nm and $\alpha = 0.027$.}
    \label{fig:rho_short}
\end{figure}

As a result, we have the smeared DOS as a function of the Fermi level
\begin{align}\label{eq:gmu_avg}
    \ev{g(E_F)} = \int_{-\infty}^{\infty} d\mu g(\mu) \frac{\exp[-\frac{(\mu - E_F)^2}{2\Gamma^2(E_F)}]}{\sqrt{2\pi \Gamma^2(E_F)}},
\end{align}
as shown in Fig.~\ref{fig:tdos}.
Replacing $g(\mu)$ by $\ev{g(E_F)}$ in Eqs.~\eqref{eq:rate_short} and \eqref{eq:rho_short}, and replacing $\abs{\mu}$ by $\sqrt{E_F^2 + 2\Gamma(E_F)^2/\pi}$ in Eq.~\eqref{eq:conduct}, we get the wire resistance as a function of the Fermi level, as shown in Fig.~\ref{fig:rho_short}.

The relatively weak dependence of TI wire  resistance on $E_F$ shown in Fig.~\ref{fig:rho_short} agrees with the experiments~\cite{munning2021}. 
However, small subband-related oscillations of the resistance seen in the experiment show up in our theory only at $N$ substantially smaller than $10^{18}$ cm$^{-3}$.

\section{Conclusion}

In the above calculations, we deal with a TI wire in a low-dielectric-constant environment without attached superconductors, and showed that at $N = 10^{19}$ cm$^{-3}$ the amplitude of potential fluctuations $\Gamma \sim 3\Delta$. In fact, a future MBS device will contain a superconductor stripe on the top of the TI wire, which may substantially screen the potential and reduce $\Gamma$. 

Indeed, the superconductor stripe provides an image charge to each impurity inside the wire, and at $\abs{x} > a$
the resulting dipole potential is $\sim e^2 a^2/\kappa \abs{x}^3$ and decays very fast. Therefore, we need to consider the contribution of impurities only at distances $r \lesssim a$ from a surface observation point. They provide normal Coulomb potential because screening by surface electrons or holes at such small distances can be ignored. 
In the case $Na^3 \gg 1$ we have
\begin{align}\label{eq:metal_gamma}
    \Gamma^2 = \beta Na^3 (e^2/\kappa a)^2,
\end{align}
where $\beta$ is an unknown numerical coefficient that depends on geometries of the wire cross-section and of the adjacent superconductor.
A similar result can be obtained by replacing $r_s$ with $a$ in Eq.~\eqref{eq:phi2}. 
For $N = 10^{19}$ cm$^{-3}$, $\kappa = 200$, $a = 20$ nm, and for a not completely unreasonable $\beta = 4\pi$, we get $\Gamma = 12$ meV, close to $\Delta = 13$ meV.

However, this $\Gamma$ is not small enough for a MBS device, which has to keep the chemical potential inside the same subband gap in order to avoid accidental Majorana pairs. 
For a long wire, even a single rare segment of the length $2a$ with a potential larger than $\Delta/2$ is detrimental. 

Let us estimate the probability of keeping the chemical potential within the same subband gap along the whole wire. 
For that we divide the long wire into $L_x/2a$ pieces with the length $2a$ each. 
For randomly positioned Coulomb impurities, the potential of each piece is an independent random variable following Gaussian distribution, and the probability for a single piece to fail is 
\begin{align}\label{eq:p}
    p = 1-\int_{-\Delta/2}^{\Delta/2} du\frac{e^{-u^2/2\Gamma^2}}{\sqrt{2\pi} \Gamma} = {\rm erfc}(\Delta/2\sqrt{2} \Gamma),
\end{align}
where ${\rm erfc}(x)$ is the complementary error function.
The probability for the whole wire to function is given by
\begin{align}\label{eq:pt}
    p_w = (1-p)^{L_x/2a}.
\end{align}
For example, in order to have $p_w \geq 0.5$ for a typical wire with $L_x = 1$ $\mu$m and  $a=20$ nm we need $\Gamma/\Delta \leq 0.23$~\footnote{In this case Eq.~\eqref{eq:p} gives $p = 0.03 \ll 1$, so that Eq.~\eqref{eq:pt} can be written as $p_w \approx \exp(-p L_x/2a)$.}.
Using Eq.~\eqref{eq:metal_gamma} with $\beta=4\pi$ we find necessary impurity concentration $N \leq N_c = 7 \times 10^{17}$ cm$^{-3}$.

To make disorder effects even weaker one can put MBS devices on high-dielectric-constant substrate, say, insulating STO~\cite{jauregui2016}. 
In this case, one can expect a 50$\%$ yield of functioning MBS devices at a few times larger $N_c$.

Thus, we find that necessary reduction of the concentration of Coulomb impurities in TI nanowires to make functioning MBS devices is similar to that in hybrid Rashba nanowires~\cite{woods2021}. In both cases, the concentration of Coulomb impurities should be reduced at least by 10 times.

\medskip

\begin{acknowledgments}
We are grateful to Y. Ando, A. Kamenev, H. Legg, V. Pribiag, A. Rosch, and B. Skinner for reading the paper and useful comments.
Y.H. was partially supported by the William I. Fine Theoretical Physics Institute.
\end{acknowledgments}

%

\end{document}